\documentclass[manuscript]{aastex}

\newcommand\eez[1]{\mbox{$10^{#1}$}}            % just 10^#2
\newcommand\hi{\ion{H}{1}}

\newcommand\civ{\ion{C}{4}}
\newcommand\ovi{\ion{O}{6}}
\newcommand\mgii{\ion{Mg}{2}}
\newcommand\mgi{\ion{Mg}{1}}
\newcommand\fei{\ion{Fe}{1}}
\newcommand\feii{\ion{Fe}{2}}
\newcommand\SII{\ion{S}{2}}
\newcommand\SIII{\ion{S}{3}}

\newcommand\lya{\mbox{Ly$\alpha$}}
\newcommand\lyb{\mbox{Ly$\beta$}}

\newcommand\msun{\hbox{$M_{\odot}$}}

\newcommand\V{\mbox{$V$}}%% V filter

\newcommand\kps{\mbox{${\rm km~s^{-1}}$}}
\newcommand\cm{\mbox{${\rm cm^{-2}}$}}

\newcommand{\bc}{\begin{center}}
\newcommand{\ec}{\end{center}}
\def\rahour{\mbox{$^{\mathrm h}$}}%
\def\ramin{\mbox{$^{\mathrm m}$}}%

\slugcomment{Submitted to the Astrophysical Journal}

\shorttitle{}
\shortauthors{Bregman et al.}

\begin{document}

\title{Outflow vs.~Infall in Spiral Galaxies: Metal Absorption in the Halo
of NGC 891}

\author{Joel N. Bregman\altaffilmark{1}, Eric D. Miller\altaffilmark{2}, 
Patrick Seitzer\altaffilmark{1}, C.R. Cowley\altaffilmark{1}, and
Matthew J. Miller\altaffilmark{1} }

\altaffiltext{1}{Department of Astronomy, University of Michigan, 
Ann Arbor, MI 48105; jbregman@umich.edu}
\altaffiltext{2}{Kavli Institute for Astrophysics and Space Research,
Massachusetts Institute of Technology, Cambridge, MA 02139}

\begin{abstract}
Gas accreting onto a galaxy will be of low metallicity while halo gas due
to a galactic fountain will be of near-solar metallicity.  We test these
predictions by measuring the metal absorption line properties of halo gas 5
kpc above the plane of the edge-on galaxy NGC 891, using observations taken
with HST/STIS toward a bright background quasar.  Metal absorption lines
of \feii, \mgii\ and \mgi\ in the halo of NGC 891 are clearly seen, and
when combined with recent deep \hi\ observations, we are able to place
constraints on the metallicity of the halo gas for the first time.  The
\hi\ line width defines the line broadening, from which we model opacity
effects in these metal lines, assuming the absorbing gas is continuously 
distributed in the halo.  The gas-phase metallicities are 
[Fe/H] = $-1.18 \pm 0.07$ and [Mg/H] = $-0.23 ~+0.36/-0.27$ (statistical errors)
and this difference is probably due to differential depletion onto grains.
When corrected for such depletion using Galactic gas as a guide, both
elements have approximately solar or even supersolar abundances.  
This suggests that the gas is from the galaxy disk, probably expelled 
into the halo by a galactic fountain, rather than from accretion of intergalactic 
gas, which would have a low metallicity.
The abundances would be raised by significant amounts if the absorbing gas
lies in a few clouds with thermal widths smaller than the rotational 
velocity of the halo.  If this is the case, both the abundances and [Mg/Fe]
would be supersolar.
\end{abstract}


%%%%%%%%%%%%%%%%%%%%%%%%%%%%%%%%%%%%%%%%%%%%%%%%%%%%%%%%%%%%%%%%%%%%%%%%
% SECTION -- Introduction
%%%%%%%%%%%%%%%%%%%%%%%%%%%%%%%%%%%%%%%%%%%%%%%%%%%%%%%%%%%%%%%%%%%%%%%%
\section{INTRODUCTION}
\label{sect:intro}

The formation of galactic disks occurs because gas with net angular
momentum is accreted into a potential well before forming stars (e.g., 
review by \citealt{silk2012}.  The
accretion can occur in two forms, where the gas is either at the
temperature of the potential well (hot), or well below that temperature.
If the gas is hot, radiative cooling leads to the creation of cool gas near
the bottom of the potential well (the cooling flow model).  Alternatively,
the gas might already be cool (e.g., \hi) falling nearly ballistically onto
the disk (cold flows), where it probably shocks and radiatively cools
before settling down onto the disk.

One or perhaps both of these processes were active when the galaxy disk
formed and most of the disk stars were produced, typically 5--12 Gyr ago.
In addition, accretion may be going on today in local galaxies at modest
accretion rates.  Gathering observational evidence for accretion today is
of considerable importance since understanding accretion today can provide
insights into the physics surrounding disk formation and evolution. 

Both hot and cold gas is seen in the vicinity or extremities of galaxies
and it could be accretion material. Examples of this are the high velocity
clouds of neutral hydrogen around the Milky Way \citep{WvW97} or M31 \citep{west2005} 
and the hot X-ray emitting atmospheres around edge-on spiral galaxies
\citep{BregmanPildis1994,Stricklandetal2004,Tullmannetal2006,Li2012}.  The
X-ray emitting halos of galaxies are largely related to the degree and
extent of star formation in the disk, so this component is most likely due
to a galactic fountain \citep{Tullmannetal2006}.  The situation with neutral
gas likely points to at least two components: a galactic fountain origin
for the intermediate velocity clouds (and some high velocity clouds) of
\hi; and the accretion of extragalactic \hi\ clouds, either from nearby
dwarf companions (e.g., the Magellanic Stream from the LMC and SMC) or
through condensation out of an extensive, dilute hot halo (much larger than
10 kpc and as yet unobserved in X-rays; \citealt{peek2008, Milleretal2009}
and references therein).

\begin{figure}
\includegraphics[width=.5\linewidth]{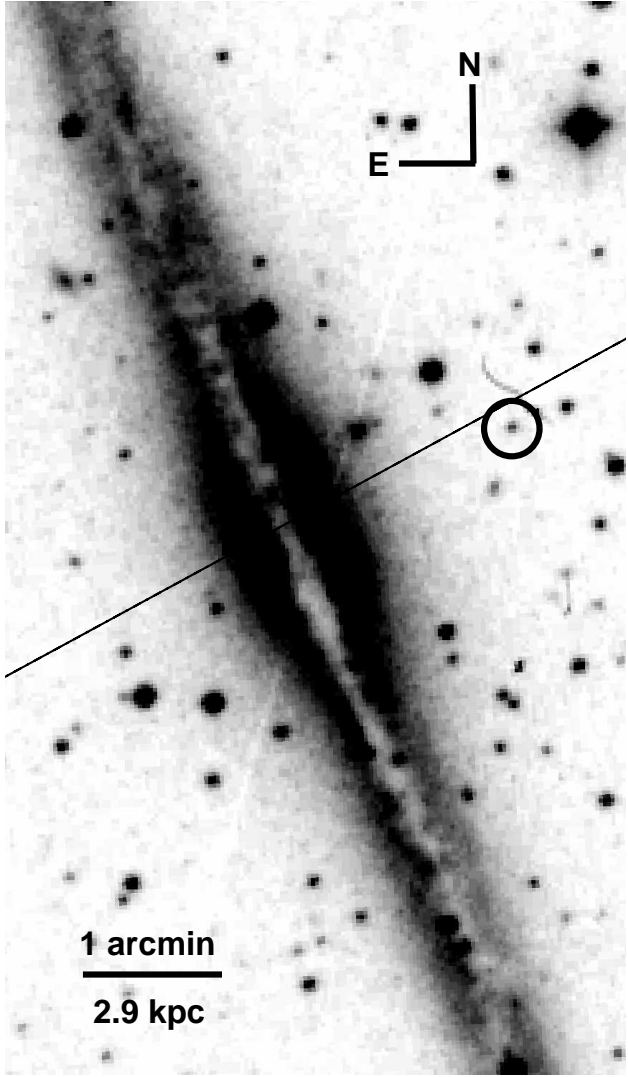}
\caption{The edge-on galaxy NGC 891 with the background AGN marked with 
a circle.  The AGN lies 106\arcsec\ (5.1 kpc) from the midplane and 
8\arcsec\ (340 pc) from the minor axis, which is estimated from the \hi\
data of \citet{Oosterlooetal2007} and marked here by a line.}
\label{fig:agn}
\end{figure}

Metallicity can provide insights into these processes since galactic gas 
will be enriched, gas torn from dwarf galaxies is of lower metallicity, 
while condensations from a large dilute halo may be close to primordial.  
Provided that these components can be separated either spatially or in 
velocity space, metallicities can identify their relative importance.  
The value of this approach is indicated by absorption line Milky Way 
studies, where some halo high velocity clouds are found to have low 
metallicity (0.3 solar), demonstrating that some extragalactic gas is 
being accreted by the Milky Way \citep{wakker2004}.  
Until now, similar observations have not been possible for external 
galaxies, as it would require a background source such as an AGN to 
be projected behind the halo of an edge-on galaxy.  Exactly that 
situation exists for an AGN that lies far behind the edge-on galaxy, 
NGC 891, where the AGN is projected 5 kpc from the galactic plane in 
the inner part of the galaxy (see Figure \ref{fig:agn}).

The edge-on galaxy NGC 891 is similar to the Milky Way in galaxy type and
optical luminosity and it lies close enough (10 Mpc) that it has been
mapped extensively in several ISM components, including \hi, X-ray emitting
gas, and warm ionized gas
\citep{RandKulkarniHester90,BregmanPildis1994,Rand97,SwatersSancisivanderHulst97,Healdetal2006}.
A heroic 21 cm observation \citep{Oosterlooetal2007}
is deep enough to map low column densities to heights as great as 22
kpc above the plane, and \eez9 \msun\ of \hi\ gas lies in the halo.  They argue
that the \hi\ halo is a combination of infalling material and a galactic
fountain.  Their \hi\ map covers the region of the AGN, so the \hi\ column
density along that sightline is known.  Their HI study leads to an essential
improvement in our original analysis \citep{miller2000}, since, by 
measuring the column density of metal lines, we can now determine
the metallicity of the halo gas.  In the following sections, we present the HST spectroscopic
observations, the resulting column density, and the metallicity implied.  We
conclude with a summary and interpretation of these results.

%%%%%%%%%%%%%%%%%%%%%%%%%%%%%%%%%%%%%%%%%%%%%%%%%%%%%%%%%%%%%%%%%%%%%%%%
% SECTION -- Data
%%%%%%%%%%%%%%%%%%%%%%%%%%%%%%%%%%%%%%%%%%%%%%%%%%%%%%%%%%%%%%%%%%%%%%%%
\section{OBSERVATIONS \& DATA ANALYSIS}
\label{sect:data}

The background AGN that we observed with HST/STIS was identified as an
X-ray source in a ROSAT image at J2000 coordinates RA,Dec =
02\rahour22\ramin24\fs45,$+$42\arcdeg21\arcmin38\farcs8.  
The discovery technique, described in \citet{KnezekBregman1998}, 
used follow-up optical spectroscopy with the MDM 2.4m telescope.  
A series of five 600 second exposures was obtained using the Modular
Spectrograph on the MDM 2.4-m Hiltner telescope during fall 1996.  The 85-mm camera with
a 600l/mm grating was used in first order with a thinned, backside
illuminated SITe CCD.  The sampling was 4.7 Angstoms per pixel.  Only
the wavelength range of 3800 - 7600 Angstroms is considered to prevent
effects from overlapping spectral orders.
The spectrophotometric standard star Feige 34 was used to flux calibrate
the combined spectrum.  However, the night was not photometric and thus
only relative fluxes are available.

\begin{figure}
\includegraphics[height=\linewidth,angle=270]{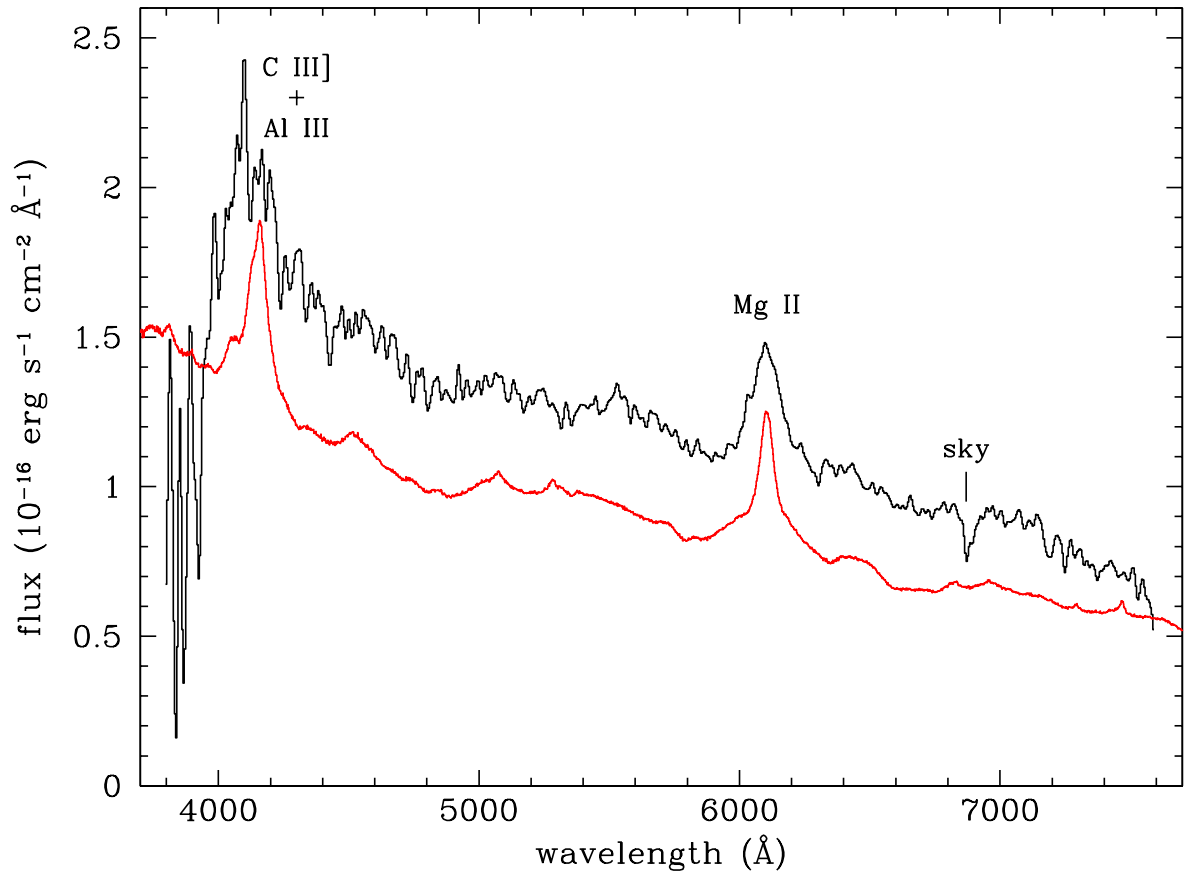}
\caption{Optical spectrum (black) of the quasar taken with the MDM 2.4m
telescope.  The SDSS composite quasar spectrum
\citep[red;][]{Vandenberketal2001} is shown for comparison, redshifted to
$z = 1.18$.} \label{fig:mdmspec} \end{figure}

These spectroscopic observations revealed this X-ray source
near NGC 891 to be an AGN that is projected upon the galaxy's halo only
106\arcsec\ (5.0 kpc) from the disk and in the inner part of the galaxy,
11\arcsec\ along the disk direction ($r = 0.5$ kpc); it has \V\ = 18.7 and 
$z = 1.18$ \citep{miller2000}.  The optical discovery spectrum is shown in Figure
\ref{fig:mdmspec}, which clearly identifies this as a broad-line quasar.  
This source was also identified by the Serendipitous Extragalactic X-Ray
Source Identification program as CXOSEXSI J022224.3+422139
\citep{Harrisonetal2003}, and follow-up optical spectroscopy in that
program confirms our redshift determination and classification
\citep{Eckartetal2006}.  We note that no other nearby edge-on galaxies in
our discovery program had bright AGNs within their halos, making this a
singular target.

The metal line absorption data for this project were obtained with the STIS
spectrograph aboard HST on 6 September 2000; the QSO was observed for a
total of 13.2 ksec over the span of 10 orbits.  These observations used the
52\arcsec$\times$0.5\arcsec\ aperture, NUV-MAMA detector, and G230L grating
centered on 2376 \AA. With this setup, the data cover a usable spectral
range from 1750--3150 \AA\ with a dispersion of 1.58 \AA\ per pixel.  The
spectral resolution is 3.3 \AA\ (413 \kps) FWHM at 2400 \AA.

\begin{figure}
\includegraphics[height=\linewidth,angle=270]{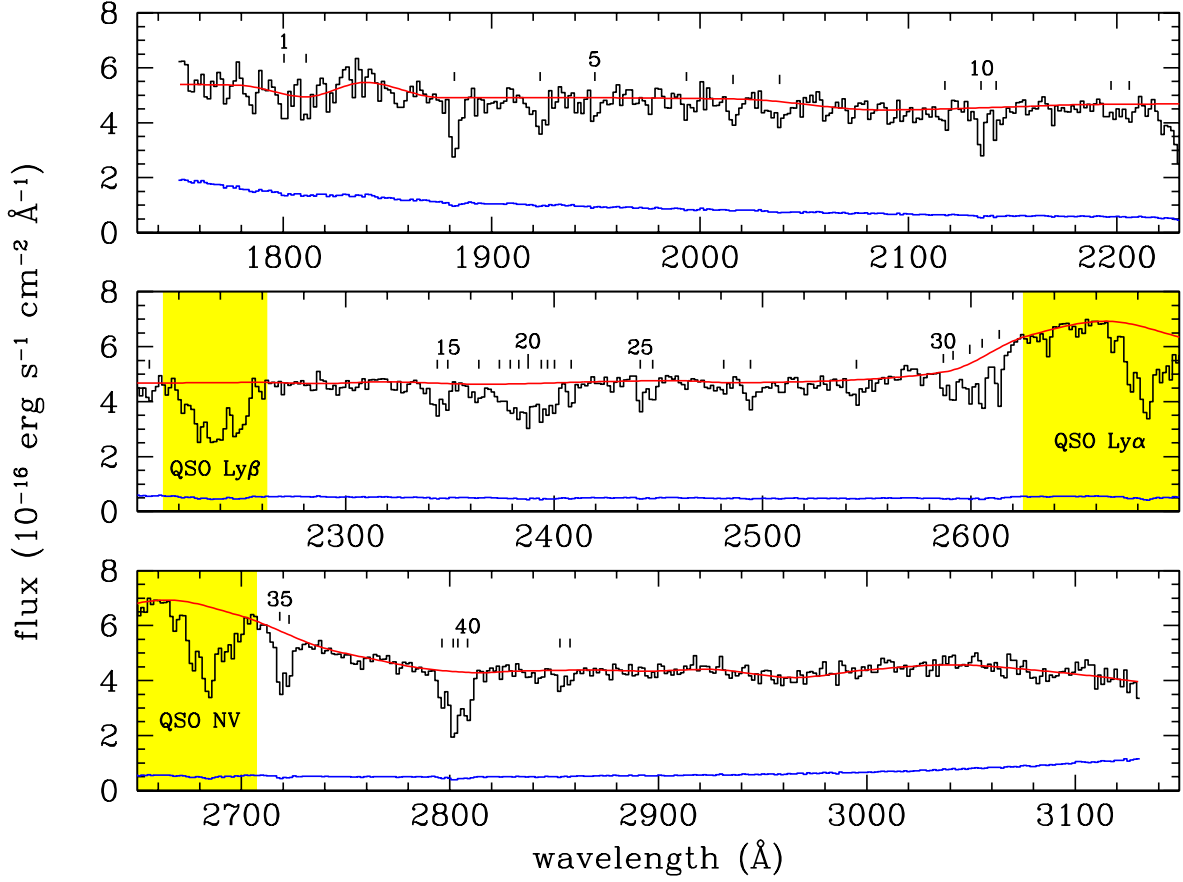}
\caption{HST STIS/NUV spectrum of the QSO behind NGC 891. The red line
shows the QSO continuum fit, the blue line shows the 3-$\sigma$
uncertainty.  Tick marks indicate detected absorption features which
correspond to entries in Table \ref{tab:lines}. The three broad absorption
regions marked in yellow near 2240, 2600, and 2680 \AA\ are due to QSO
absorption, and have been ignored in the line searching and analysis.}
\label{fig:spec}
\end{figure}

The individual exposures were scaled by the exposure time and combined.
We identified the continuum (including QSO emission lines) along the lines
described in the HST Quasar Absorption Line Key Project \citep{Schneideretal1993}. 
Figure \ref{fig:spec} shows the combined
spectrum along with the continuum fit and propagated 3-$\sigma$ errors.
The resulting spectrum was searched for absorption features using the
automated absorption line search algorithm available in the Key Project
software.  The procedure followed was identical to that described in
\citet{Milleretal2002}.  Detected absorption lines were verified by eye.
We also used SPECFIT (within STSDAS) and the fitting procedures within
Origin Pro 8.5 to extract equivalent widths, obtaining similar results.  
It is essential to use the true instrumental line shape rather than just a Gaussian,
otherwise one will obtain systematically smaller equivalent widths.

The QSO spectrum lacks the bright emission lines usually associated with
such objects, and in fact shows complex absorption in the \lyb\ and \ovi\
emission line region (near 2240 \AA).  Corresponding absorption is seen
near the very weak \lya\ emission line (2650 \AA) and the region where we
expect \ion{N}{5}\ emission (2700 \AA).  Indeed, the
\ion{C}{3}]$+$\ion{Al}{3} peak in the optical spectrum (see Figure
\ref{fig:mdmspec}) also appears weaker then the composite quasar spectrum.
The \ion{Mg}{2}\ line, while fairly strong, lacks the narrow component seen
in the composite.  It is well-fit by a Gaussian with FWHM $= 136.3 \pm 4.7$
\AA\ ($6680 \pm 230$ \kps).
A detailed QSO classification requires additional spectroscopic analysis
from a broader UV and optical band, and since we are only interested in the
quasar as a background light source, such analysis is beyond the scope of
the current work.
Due to the complexities of the
QSO \lya\ and \lyb\ features, we have excluded these regions from our
analysis.

A number of absorption features were found with greater than 3-$\sigma$
significance; these are marked in Figure \ref{fig:spec} and listed in Table
\ref{tab:lines} along with the best-fit line centers and equivalent widths.
Due to the poor spectral resolution, each detected line was unresolved and
we are unable to constrain the intrinsic line widths.  To identify the
absorbers, we first noted the expected Galactic and NGC 891 lines.  We
clearly see the strong \mgii\ $\lambda\lambda$2796,2803 doublet from NGC
891, although it is partially blended with its Galactic counterpart.
The \feii\ $\lambda$2344 line is detected in both the Galaxy and NGC 891 without 
contamination. 
Unfortunately, there is severe blending of the \feii\ $\lambda\lambda$ 2374,
2383, likely with \lya\ forest lines.  The best-fit decomposition is given 
in Table \ref{tab:lines} (17-24) but we cannot obtain unique equivalent widths 
for these two lines at above 3-$\sigma$, so they are not used in the determination
of the \feii\ column density.
Similarly, line blending in 2570--2610 \AA\, while less severe, caused unacceptably large systematic 
uncertainties for the \feii\ $\lambda\lambda$ 2587, 2600 lines, so the only
line used for the analysis of the \feii\ column is the $\lambda$2344 line.
We note that had we used these other lines, a larger \feii\ column density and metallicity
would have resulted.

The remaining lines did not match any expected metal resonance absorption
lines from the Galaxy or NGC 891, and they were taken to be intervening
absorption systems.  A simple matching procedure was implemented to
identify single systems with multiple features.  We discover a strong
absorber at $z = 0.756$ with lines from \lya, \lyb, \ovi\ $\lambda$1032,
\ion{Si}{3}\ $\lambda$1206, \ion{Si}{4}\ $\lambda\lambda$1393,1402, and
\civ\ $\lambda\lambda$1548,1550.  The measured equivalent widths of these
species are similar to those for other \civ\ absorption systems 
identified in the literature \citep[e.g.,][]{Sargentetal1988}.
Remaining absorption lines were taken to be \lyb\ if a corresponding \lya\
feature at the same redshift could be identified, otherwise they were
assumed to be \lya\ from intervening clouds.  These results are included in
Table \ref{tab:lines} for completeness, but they will not be discussed
further in the present work.
% In the table, have a footnote for badly blended lines

The wavelengths of the detected Galactic lines were not consistent with
zero velocity, indicating a wavelength zeropoint offset likely due to the
source not being centered in the STIS aperture.  We estimated an offset of
$-1.6 \pm 0.6$ \AA\ from the best-fit centers of the clean Galactic
absorption lines, assuming they were at $v_r = 0$ \kps.  This corresponds
to about 1 pixel or 0.025\arcsec, larger than the nominal pointing error of
0.01\arcsec\ (with no source acquisition peak-up), but not uncommon.
We adjusted the spectrum and measured line centers to correct for this
offset; those changes are reflected in Table \ref{tab:lines} and Figure
\ref{fig:spec}.  The lines from NGC 891 are largely consistent with the
measured systemic velocity of $+528$ \kps, which is also close to the
measured \hi\ velocity at this location above the disk
\citep{Oosterlooetal2007}.  

\begin{figure}
\includegraphics[width=\linewidth]{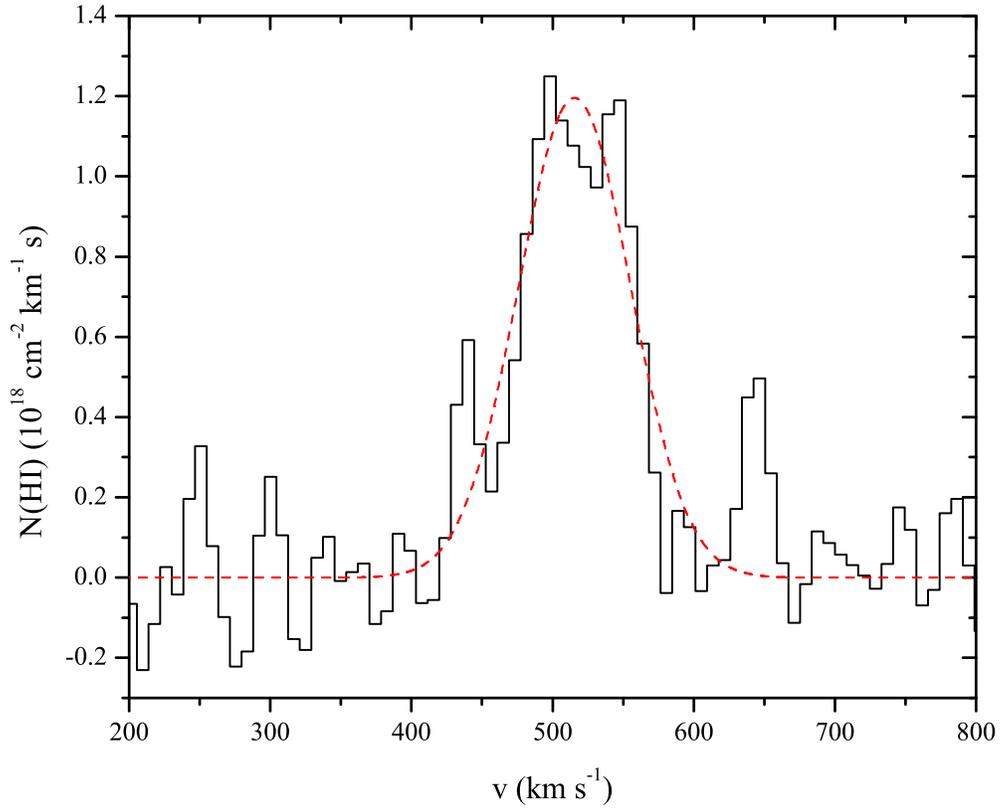}
\caption{\hi\ 21 cm emission spectrum through the halo of NGC 891, in the
direction of the QSO.  The vertical axis shows \hi\ column density per
\kps.  The red dashed line traces the best-fit Gaussian profile, with a
width of $93 \pm 8$ \kps\ FWHM.  Data are from \citet{Oosterlooetal2007}.}
\label{fig:hispec}
\end{figure}

Clean detections of \feii\ $\lambda$2344, and \mgii\ $\lambda$2803 from 
the halo of NGC 891 allow us to estimate column densities for these species.
The \mgi\ $\lambda$2853 absorption line is detected at 5$\sigma$ at the Milky Way 
velocity but only at 3$\sigma$ from NGC 891, so we treat it as an upper limit.
To obtain column densities requires corrections for the optical depth of the lines.
There are not enough lines to fit a curve of growth for any one ion, from
which a Doppler broadening parameter would be obtained.  Fortunately, these
ions and HI should be cospatial and the 
Doppler broadening parameter is available directly from the 21 cm \hi\ 
observations, which are optically thin and have excellent velocity 
resolution.  From the data that were obtained by \citet{Oosterlooetal2007},
an \hi\ spectrum was extracted at the location of the background AGN (see
Figure \ref{fig:hispec}).
The spectrum shows a distribution that is typical of a rotating disk, having
a flat-top profile with modest double-horns.  To simplify matters, we fit
the profile with a Gaussian where the FWHM = 93 $\pm$ 8 \kps.  From the 
zero moment map, the total \hi\ column along this line of sight is 
1.3$\times$10$^{20}$ \cm.  With this column and line width, the optical
depth at line center may be quite significant, depending upon the elemental
abundances, since depletion on to grains may be important.

To infer the column densities for \feii\ $\lambda$2344, and \mgii\ $\lambda$2803, 
we need to correct for optical depth effects, along with the ionization fraction.
When the optical depth is small, the conversion is linear and independent of 
the number of components.  However, for significant optical depths, the result 
depends not only on the equivalent width, the f value, and the Doppler parameter, 
it depends on whether there are multiple components.  The existence of multiple 
components as well as their properties cannot be resolved with our data, so we 
proceed by assuming that the gas is uniformly distributed in the halo.  
Later, we discuss the degree by which the inferred column density underestimates 
the value if multiple components exist.  The most strongly affected line 
is \mgii\ $\lambda$2803, although the correction to \feii\ $\lambda$2344 is not negligible.  

For a single-component model, we construct a curve of growth assuming the
usual Voigt profile with the Doppler parameter taken from the \hi\ line, 
where $b = 56$ \kps.  We use the atomic data from \citet{Morton2003}
to determine oscillator strengths, statistical weights, and partition
functions for each transition (see \citealt{Spitzer78} for details of the
curve of growth method).  Statistical uncertainties in the equivalent width 
determination and uncertainty in the Doppler velocity parameter 
are propagated through to provide column density errors.  Additional uncertainties
such as those associated with various assumptions are discussed below 
(e.g., we assume a single line characterized by a single value of $b$, but there may be multiple lines).
The \feii\ $\lambda$2344 line is close to the linear part of the curve of growth,
with a correction factor to the equivalent width of 1.69.  
The resulting column for \feii\ is
$\log[N($\feii$)/\cm] = 14.41 \pm 0.14$, while the upper limit for the \mgi\
column is $\log[N($\mgi$)/\cm] = 12.67 $ (statistical errors only).  
The \mgii\ $\lambda$2803 line has an optical depth of about 70
at the line center and lies on the the flat part of the curve
of growth.  We calculate a column density of $\log[N($\mgii$)/\cm] = 15.48
(+0.36,-0.27)$.

\section{RESULTS \& DISCUSSION}

To convert the equivalent width to a column density of an element 
requires a model for both the opacity effects (discussed above) and 
for the ionization fraction for the particular ionization state.  
We assume that photoionization controls the ion fractions and that 
the photons come from the disk of the galaxy.  
This is certainly not the entire story, as \citet{Rand1998} and 
\citet{Rand2008} have shown that photoionization models have difficulty 
reproducing the optical and infrared emission line data from the halo of NGC 891.  
They conclude that there is likely to be a secondary source of ionization. 
\citet{Reynolds1999} reach a similar conclusion for Milky Way halo 
emission line gas and they suggest that the additional heating could be 
due to the dissipation of interstellar plasma turbulence. 

Even by ignoring this additional heating mechanism, the precise value 
of the ionization radiation field is only known roughly, with uncertainties 
from both models and the observational interpretation of emission lines.  
An important input to the models is the leakage of ionizing radiation from 
the disk, along with the propagation of UV radiation between the ionization 
edges of hydrogen (13.6 eV) and the ions \mgi\ (7.65 eV) and \fei\ (7.9 eV), 
which provides the dominant amount of photoionization for \mgii\ and \feii\ 
if too little of the radiation above 13.6 eV escapes.  
The propagation and escape of ionizing radiation has been calculated with 
various models (e.g., \citealt{Dove1994,Wood2010}) where the result depends 
on the structure of the ISM and the degree to which radiation propagates 
through low density channels where absorption is minimized.  
These models are constrained by observations, such as by galaxy-wide emission 
line observations (e.g., \citealt{Oey2007}) and by emission line observations 
directly from the halos of edge-on galaxies \citep{Rand1998} or from Galactic 
high velocity clouds that are irradiated \citep{Putman2003}.

We adopt a simple model for the ionizing and UV radiation reaching 5 kpc 
in which the shape of the radiation field is given by O stars and the value 
of the ionization parameter is constrained by \citet{Rand1998}.  
Calculations were performed with Cloudy version C08, most recently
described by \citet{Ferlandetal1998}.  The ionizing photons pass through
the halo \hi\ observed by \citet{Oosterlooetal2007} prior to ionizing the 
gas at 5 kpc above the disk.  
We assume that the \hi\ is smoothly distributed at any particular height so that
the plane-parallel approximation holds.
The radiation softens as it rises upward through the \hi\ layer so that by the time
it reaches 5 kpc, most of the radiation above 13.6 eV has been absorbed.
From the work of \citet{Rand1998}, the ionization parameter U is likely in the 
range -6 $<$ $log(U)$ $<$ -3.5, from which we calculate that the ionization fractions of
\mgii\ and \feii\ exceed $>$ 90\%.  
There is a simple explanation for this result in that the \hi\ absorbs most 
of the photons above 13.6 eV, while the ionization potential for the transition 
to twice ionized species is 15.0 eV for \mgii\ and 16.2 eV for \feii.  
Therefore, there are ample photons to ionize these elements to their first 
ionization state but not the second.  For this reason, this result is nearly 
independent of the flux density of radiation emitted from the disk, the 
vertical form for the structure of the \hi\ or even the total column 
of \hi\, provided they not orders of magnitude different than observed.
There may be complications, discussed below, so for our limiting case, 
we adopt ionization fractions of 1.0 for these ions.  

Emission from ionized gas is detected to heights approaching 5 kpc \citep{Rand2008}, 
indicating that the Mg and Fe are not entirely singly ionized.  Models now try to
take into account the complicated non-uniform gaseous structures that would be
produced by supernovae \citep{Wood2010}.  In their models with high ionizing fluxes, gas is nearly
fully ionized at all heights above the plane, in conflict the the HI observations showing
large amounts of HI above the disk.  For models with lower ionizing fluxes, the gas is
ionized within 1 kpc of the plane, becoming entirely neutral at greater heights, 
consistent with our limiting case model.  

Another relevant study is the absorption
line study toward the globular clusters M3 and M5 at heights of 10 kpc and 5.3 kpc above 
the plane \citep{howk12}.  They find that nearly all of the S along the sight lines
are in the form of \SII\ or \SIII\ with 50-75\% of S as \SII.  The ionization potentials
of \ion{S}{1} is 10.4 eV, below that of \hi\ but about 3 eV above that of \mgi\ or \fei.
The ionization potential for \SII\ is 23.3 eV, 7-8 eV above that of \mgii\ or \feii\, so 
we might estimate that about half of Mg or Fe is not in the form of \mgii\ or \feii\ along these sight lines.
We note that this determination of the relative ionization states of S probes 
a variety of heights from the disk up to the globular clusters, which may not be 
representative of the ionization state distribution for a sight line at a constant 
5 kpc height above the plane.  In the following, we determine metallicities for 
the limiting case (100\% of the Mg and Fe are singly ionized), acknowledging that
lower ionization fractions would lead to larger metallicities, for Mg and Fe.  

Some information is available for the ionized gas column densities at this
height in the halo from optical Balmer line emission and from X-ray
studies.  The X-ray observations show a hot halo of temperature about 
$3 \times 10^6$ K, which is close to the value expected for hydrostatic
equilibrium, so it is probably volume-filling \citep{BregmanHouck97}.  The
X-ray emission leads to a density distribution, and along the absorption
line of sight closest to the bulge, the column density is $8 \times
10^{19}$ cm$^{-2}$, less than the \hi\ column.  This gas will not produce UV
absorption lines, but it will provide an ambient pressure, which is
$p/k \sim 10^4$ K cm$^{-3}$.  This implies that $10^4$ K, Balmer-emitting
gas would have a density of $\sim 1$ cm$^{-3}$, assuming it is in pressure
equilibrium with the X-ray gas.  To reproduce the observed Balmer emission
line intensity \citep{Rand97}, the $10^4$ K gas only occupies 1\% of the
volume and has a column density thirty times less than the \hi\ column, so
its contribution to the mass can be neglected.  Similarly, the \hi,
probably at $\sim 10^2$ K, would also occupy a small volume but its mass is
measured directly and is the dominant component at this height.

We can express the abundances of Mg and Fe relative to their solar
abundances, for which we adopt a solar Fe abundance of 7.50 and a
solar Mg abundance of 7.60 \citep{Asplundetal2009}.  This leads to [Fe/H] =
$-1.18 \pm 0.14$ and [Mg/H] = $-0.23 +0.36/-0.27$.  These abundances do not
take into account the depletion of metals onto grains.  Fortunately, Fe and
Mg are found to be depeleted by different amounts along Milky Way
sightlines, with the relative depletion difference being correlated with
the total depletion \citet{SavageSembach1996}.  In all cases, the Fe
depletion is greater than that of Mg, with the difference usually a factor
of 2--10 (0.3--1.0 in the log).  For absorption lines in Milky Way
disk$+$halo gas, the average logarithmic depletion correction for Fe is
0.92 and for Mg is 0.6.  Making these corrections, [Fe/H] = -0.26 $\pm$ 0.14 and
[Mg/H] = +0.37 $+0.36/-0.27$.  If we were to use the depletion correction factors for
warm disk gas, 1.22 for Fe and 0.81 for Mg (in the log), the corrected
abundances would be [Fe/H] = 0.02 $\pm$ 0.14 and [Mg/H] = 0.58 $+0.36/-0.27$.  
In either case, the metallicity of the gas is at least half solar and possibly
above the solar value.

In this analysis, we assumed that the absorbing material is distributed continuously 
in the halo so that the line width is determined by the rotational velocity 
distribution of the halo.  If the absorbing gas is confined to a few clouds or 
sheets, with line widths that are narrow compared to the rotational velocity, 
and where the lines are not overlapping, the inferred columns for Fe and Mg 
would be raised.  The limiting case would be if the intrinsic thermal lines 
widths were so narrow that the lines lie on the damping part of the curve of 
growth, which can raise the columns by two orders of magnitude.  This would 
lead to a metallicity about two orders of magnitude in excess of the solar value, 
an unlikely result.  Also, this correction would affect Mg more than Fe, so 
even more modest corrections could lead to highly supersolar values of [Mg/Fe].  
For these reasons, we expect that our adopted model is not far from the truth, 
although the columns could be 2-3 times higher without having to adopt exotic 
models for the absorbing medium (e.g., an unmixed part of supernova ejecta).

Gas being accreted onto NGC 891 is unlikely to have near-solar abundances.
For example, the HI high velocity gas around the Milky Way known as Complex C,
thought to be accretion material, has a metallicity of 0.1-0.3 Z$_{Solar}$ \citep{tripp2003}.
Consequently, we conclude that the gas is from the disk of the galaxy and
it has been brought up to a height of 5 kpc by galactic disk processes.
This favors a galactic fountain for the origin of the metal-bearing gas.
If the disk metallicity in NGC 891 is similar to the Milky Way, this sight line
above the inner disk should have a metallicity above the solar value.

There is a considerable literature on \mgii\ absorption lines seen against 
background AGN spectra, although they tend to be at much larger impact parameters 
(e.g., \citealt{Chen2010, Kacprzak2011, Bouch2012}).  These studies also have 
significantly higher spectral resolution, which permits them to study the 
velocity properties of the lines relative to the galaxy.  
From these studies, the \mgii\ absorbing material has a higher column density 
close to the galaxy and our value is consistent with some of the other 
absorption systems close to galaxies \citep{Chen2010}.  
This absorbing gas is not simply in rotation around the galaxy and various 
scenarios are suggested, such as where some of the gas is being accreted 
along the major axis with outflow along the minor axis \citep{Kacprzak2012}.  
Present observational limitations prevent us from making such velocity 
comparisons and due to the edge-on orientation, it will never be possible to 
comment on inflow vs outflow.  However, higher spectral resolution observations 
will allow one to examine whether this gas is rotating like the \hi\ or involved in 
some different type of behavior, such as the inflow along a stream of accretion material.

There are a number of improvements that are possible with additional UV
absorption line observations.  At the resolution of these observations, the
lines in NGC 891 are not only unresolved, they are partly blended with
weaker Milky Way lines.  With higher spectral resolution, we could obtain
equivalent widths for five separate \feii\ lines, covering a range of $f$
values.  With such a set of lines, a more accurate column density is easily
obtained, especially in the event that there are multiple components in a line,
due to a non-uniform cool gas distribution.  
With a resolution of 20 km s$^{-1}$ or less, we would be able to
resolve the rotation of the galaxy and determine abundances as a function
of radial position in the galaxy.  Another important improvement would be
to accurately determine the depletion factor in the gas.  In this paper, we
inferred the absolute depletion levels by using the relative depletions of
two elements that are primarily not in the gas phase.  A better way of
measuring the depletion is to use at least one element that is primarily in
the gas phase.  Low depletions are found in the refractory elements, such
as O or S, which have UV resonance lines at wavelengths shorter than the
STIS observation presented here.  Both of these observational improvements
are possible using the COS spectrograph on HST.

\acknowledgements
The authors wish to thank Tom Oosterloo for providing the original \hi\
data.  We gratefully acknowldege partial support for this work from NASA, and
in particular, through a HST grant as well as a LTSA and ADAP grant.

\clearpage

\begin{deluxetable}{lccccc}
\tabletypesize{\footnotesize}
\tablewidth{0pt}
\tablecaption{Discrete AVCs: General Properties
     \label{tab:lines}}
\tablehead{
\colhead{ID\tablenotemark{a}} &
\colhead{$\lambda_{\rm obs} \pm\sigma$ (\AA)} &
\colhead{$W_{\lambda} \pm\sigma$ (\AA)} &
\colhead{species} &
\colhead{$z$ or $v$\tablenotemark{b}} &
\colhead{system} \\
}
\startdata
1  & 1800.38 $\pm$0.75 & 0.98 $\pm$0.40 & Ly$\beta$                 & 0.756       & \civ\ absorber \\
2  & 1810.99 $\pm$0.78 & 0.88 $\pm$0.41 & \ion{O}{6} $\lambda$1032  & 0.756       & \civ\ absorber    \\
3  & 1882.23 $\pm$0.32 & 2.56 $\pm$0.30 & Ly$\alpha$                & 0.548       & \nodata   \\
4  & 1923.26 $\pm$0.63 & 1.62 $\pm$0.29 & Ly$\alpha$                & 0.582       & \nodata   \\
5  & 1949.49 $\pm$0.82 & 0.85 $\pm$0.28 & Ly$\alpha$                & 0.604       & \nodata   \\
6  & 1993.44 $\pm$0.70 & 0.68 $\pm$0.25 & Ly$\beta$                 & 0.945       & \nodata   \\
7  & 2015.92 $\pm$0.59 & 1.11 $\pm$0.24 & Ly$\beta$                 & 0.969       & \nodata   \\
8  & 2038.19 $\pm$0.56 & 1.05 $\pm$0.23 & Ly$\alpha$                & 0.677       & \nodata   \\
9  & 2117.58 $\pm$0.64 & 0.75 $\pm$0.21 & \ion{Si}{3} $\lambda$1207 & 0.756       & \civ\ absorber    \\
10 & 2134.82 $\pm$0.22 & 1.94 $\pm$0.18 & Ly$\alpha$                & 0.756       & \civ\ absorber    \\
11 & 2142.02 $\pm$0.35 & 1.13 $\pm$0.19 & Ly$\beta$                 & 1.094       & \nodata   \\
12 & 2197.10 $\pm$0.33 & 0.62 $\pm$0.18 & Ly$\alpha$                & 0.807       & \nodata   \\
13 & 2205.94 $\pm$0.58 & 0.77 $\pm$0.18 & Ly$\beta$                 & 1.150       & \nodata   \\
14 & 2344.08 $\pm$0.26 & 1.35 $\pm$0.14 & \feii $\lambda$2344       & $-17\pm33$  & MW \\
15 & 2348.98 $\pm$0.83 & 0.83 $\pm$0.14 & \feii $\lambda$2344      & $+609\pm106$ & NGC 891 \\
16 & 2363.89 $\pm$0.74 & 0.80 $\pm$0.16 & Ly$\alpha$                & 0.945       & \nodata   \\
17 & 2373.85 $\pm$0.53 & 0.81 $\pm$0.16$^B$ & \feii $\lambda$2374       & $-76\pm66$  & MW \\
18 & 2379.10 $\pm$0.38 & 1.08 $\pm$0.17$^B$ & \feii $\lambda$2374       & $+585\pm47$ & NGC 891 \\
19 & 2383.14 $\pm$0.18 & 1.11 $\pm$0.17$^B$ & \feii $\lambda$2383       & $+47\pm22$  & MW \\
20 & 2387.52 $\pm$0.40 & 1.64 $\pm$0.15$^B$ & \feii $\lambda$2383       & $+598\pm50$  & NGC 891\\
21 & 2393.94 $\pm$0.15 & 1.38 $\pm$0.16 & Ly$\alpha$                & 0.969$^b$       & \nodata   \\
22 & 2396.87 $\pm$1.05 & 0.60 $\pm$0.20 & Ly$\alpha$                & 0.972$^b$       & \nodata   \\
23 & 2400.33 $\pm$0.34 & 1.07 $\pm$0.24 & Ly$\alpha$                & 0.974$^b$       & \nodata   \\
24 & 2408.24 $\pm$0.30 & 0.97 $\pm$0.15 & Ly$\alpha$                & 0.981       & \nodata   \\
25 & 2441.44 $\pm$0.32 & 1.03 $\pm$0.15 & Ly$\alpha$                & 1.008       & \nodata   \\
26 & 2447.25 $\pm$0.39 & 0.78 $\pm$0.15 & \ion{Si}{4} $\lambda$1394 & 0.756       & \civ\ absorber    \\
27 & 2481.41 $\pm$0.88 & 0.59 $\pm$0.16 & Ly$\alpha$                & 1.041       & \nodata   \\
28 & 2494.33 $\pm$0.11 & 1.12 $\pm$0.14 & Ly$\alpha$                & 1.052       & \nodata   \\
29 & 2545.15 $\pm$0.41 & 1.03 $\pm$0.14 & Ly$\alpha$                & 1.094       & \nodata   \\
30 & 2586.68 $\pm$0.42 & 0.67 $\pm$0.14$^B$ & \feii $\lambda$2587       & $+3\pm49$   & MW \\
31 & 2591.48 $\pm$0.29 & 1.19 $\pm$0.14$^B$ & \feii $\lambda$2587       & $+560\pm33$ & NGC 891       \\
32 & 2599.39 $\pm$0.25 & 1.41 $\pm$0.13$^B$ & \feii $\lambda$2600       & $-90\pm28$  & MW \\
33 & 2605.38 $\pm$0.32 & 1.87 $\pm$0.12$^B$ & \feii $\lambda$2600       & $+600\pm36$ & NGC 891       \\
34 & 2613.48 $\pm$0.17 & 1.75 $\pm$0.12 & Ly$\alpha$                & 1.150       & \nodata    \\
35 & 2718.44 $\pm$0.15 & 2.01 $\pm$0.13 & \civ  $\lambda$1548       & 0.756       & \civ\ absorber \\
36 & 2722.93 $\pm$0.21 & 1.56 $\pm$0.13 & \civ  $\lambda$1551       & 0.756       & \civ\ absorber \\
37 & 2796.12 $\pm$0.24 & 1.56 $\pm$0.16 & \mgii $\lambda$2796       & $-24\pm25$  & MW \\
38 & 2801.57 $\pm$0.28 & 2.31 $\pm$0.61 & \mgii $\lambda$2796       & $+559\pm30$ & NGC 891 \\
39 & 2803.79 $\pm$0.80 & 1.26 $\pm$0.58 & \mgii $\lambda$2804       & $+27\pm86$  & MW \\
40 & 2808.57 $\pm$0.20 & 2.28 $\pm$0.16 & \mgii $\lambda$2804       & $+538\pm21$ & NGC 891 \\
41 & 2852.81 $\pm$0.47 & 0.83 $\pm$0.17 & \mgi  $\lambda$2853       & $-15\pm49$  & MW \\
42 & 2857.59 $\pm$0.76 & 0.50 $\pm$0.17 & \mgi  $\lambda$2853       & $+485\pm79$ & NGC 891 \\
\enddata
\tablenotetext{a}{ID values correspond to the lines marked in Figure
\ref{fig:spec}.}
\tablenotetext{b}{Values with $\pm$ 1-$\sigma$ errors indicate velocities
in \kps; other values are redshifts.}
\tablenotetext{B}{Indicates blending that makes the EW values not sufficiently reliable for analysis.}
\end{deluxetable}

\clearpage

\end{document}